\documentclass[nohyperref]{article}

\usepackage{microtype}
\usepackage{graphicx}
\usepackage{subfigure}
\usepackage{booktabs} 

\usepackage{hyperref}

\usepackage[accepted]{icml2022}

\usepackage{amsmath}
\usepackage{amssymb}
\usepackage{mathtools}
\usepackage{amsthm}
\usepackage{float}

\usepackage[capitalize,noabbrev]{cleveref}

\theoremstyle{plain}

\theoremstyle{definition}

\theoremstyle{remark}

\usepackage[textsize=tiny]{todonotes}

\icmltitlerunning{Neural Posterior Estimation with Differentiable Simulators}

\begin{document}

\twocolumn[
\icmltitle{Neural Posterior Estimation with Differentiable Simulators}



\icmlsetsymbol{equal}{*}

\begin{icmlauthorlist}
\icmlauthor{Justine Zeghal}{equal,apc}
\icmlauthor{François Lanusse}{equal,aim}
\icmlauthor{Alexandre Boucaud}{equal,apc}
\icmlauthor{Benjamin Remy}{aim}
\icmlauthor{Eric Aubourg}{apccea}
\end{icmlauthorlist}

\icmlaffiliation{apc}{Université Paris Cité, CNRS, Astroparticule et Cosmologie, F-75013 Paris, France}
\icmlaffiliation{aim}{Université Paris-Saclay, Université Paris Cité, CEA, CNRS, AIM, 91191, Gif-sur-Yvette, France}
\icmlaffiliation{apccea}{Université Paris Cité, CNRS, CEA, Astroparticule et Cosmologie, F-75013 Paris, France}

\icmlcorrespondingauthor{Justine Zeghal}{zeghal@apc.in2p3.fr}

\icmlkeywords{Machine Learning, ICML, SBI}

\vskip 0.3in
]



\printAffiliationsAndNotice{\icmlEqualContribution} 

\begin{abstract}
Simulation-Based Inference (SBI) is a promising Bayesian inference framework that alleviates the need for analytic likelihoods to estimate posterior distributions. Recent advances using neural density estimators in SBI algorithms have demonstrated the ability to achieve high-fidelity posteriors, at the expense of a large number of simulations ; which makes their application potentially very time-consuming when using complex physical simulations. In this work we focus on boosting the sample-efficiency of posterior density estimation using the gradients of the simulator. We present a new method to perform Neural Posterior Estimation (NPE) with a differentiable simulator. We demonstrate how gradient information helps constrain the shape of the posterior and improves sample-efficiency.

\end{abstract}

\section{Introduction}
\label{sec:introduction}
Various scientific fields use complex computer simulations to describe real physical processes as accurately as possible. These simulators map the parameter space $\theta$ to a simulation x through an implicit likelihood $p(\text{x}|\theta)$ that makes the inverse process of constraining the parameter space from data intractable in practice. Simulation-Based Inference (SBI), also known as Likelihood-free inference, provides a framework to alleviate that problem in different ways.

One can approximate the likelihood distribution $p(\text{x}|\theta)$  \citep{wood_statistical_2010,nle1,nle2}, or the  likelihood ratio $r(\text{x}, \theta', \theta_t) = \frac{p(\text{x}|\theta')}{p(\text{x}|\theta_t)}$ from the simulations \citep{lr1,lr2,lr3,lr4}, and use a sampling method to estimate the posterior. Others choose an amortized method by directly 
approximating the posterior distribution $p(\theta|\text{x})$ \citep{pe,npe1,npe2,npe3}.
But such methods treat the simulator as a black-box implicit distribution by considering only forward simulations and discarding all information on the internal process.

\citet{mining_gold} proposed to work with augmented data such as the gradients of the simulator and introduced a way to approximate the likelihood distribution and the likelihood ratio which leverages this augmented data and thus improves sample efficiency and inference quality.

In this work, we extend the work of \citet{mining_gold} and propose the first Neural Posterior Estimation method augmented with gradients of the simulator. Implementing this approach necessitates the use of a particular kind of Normalizing Flows, called Smooth Normalizing Flows \citep{smooth_nf}, which have the property of having well defined, smooth, and expressive gradients. We apply our approach on a standard SBI benchmark problem (Lotka-Volterra) and recover acceleration factors when using simulation gradients consistent with the results obtained with the NLE and NRE methods of \citet{mining_gold}.

\section{Simulation-Based Inference with Differentiable Simulators}
\label{sec:sbi}
Bayesian inference aims to infer the parameters $\theta_0$ that have generated a given observation $\text{x}_0$. From Bayes theorem we have
\begin{equation}
p(\theta|\text{x}_0) = \frac{p(\text{x}_0|\theta)p(\theta)}{p(\text{x}_0)} \propto p(\text{x}_0|\theta)p(\theta)\,,
\end{equation}
and since $\text{x}_0$ is the result of a large number of transformations involving a large number of latent variables $z$, the marginal likelihood
\begin{equation}
p(\text{x}_0|\theta) = \int p(\text{x}_0|\theta, z)p(z)dz\,,
\end{equation}
is intractable. 

SBI is particularly useful in this case since it provides a framework to approximate the posterior without using an analytic likelihood. To approximate $p(\theta|\text{x}_0)$ SBI algorithms require the observation $\text{x}_0$, a prior $p(\theta)$ for the model parameters $\theta$ and a simulator $\text{x} \sim p(\text{x}|\theta)$ to sample from the intractable likelihood.

In this work we are interested in doing NPE which aims to directly learn the following distribution 
\begin{equation}
    p(\theta | \text{x}) \propto \int p(\text{x} | \theta, z) p(z|\theta) p(\theta) dz\,,
\end{equation}
using a NDE such as a Normalizing Flow (hereafter NF).
One can train the NDE to learn the approximated distribution $p_{\varphi} (\theta | \text{x})$ from samples $(\text{x}, \theta)$ of the joint distribution by minimizing
\begin{equation}
    \mathcal{L}_{\rm NLL} = \mathbb{E}_{p(\theta, \text{x})} \left[ - \log p_{\varphi} (\theta | \text{x})  \right],
    \label{eq:nll}
\end{equation}
and then evaluate it on the given observation $\text{x}_0$ to get the approximated posterior $p_{\varphi} (\theta | \text{x} = \text{x}_0 ) \approx p (\theta | \text{x}_0 )$.

As recognized in \citet{mining_gold}, when the simulators are differentiable one can extract for each simulation the gradients w.r.t. simulation parameters, which provides significantly more information than the samples from the simulator, and can be used to help constrain the posterior density estimates obtained by SBI. Therefore, in our case, we extract for each simulation $(\text{x}_i,\theta_i)$ the gradient of the joint log-probability of the simulator with respect to input parameters. In the following, we will note this gradient as
\begin{equation}
\begin{array}{ll}
\nabla_{\theta} \log p(\theta|\text{x},z) &= \nabla_{\theta}  \log p(\text{x}|\theta,z) \\
&+ \nabla_{\theta} \log p(z|\theta)\\ 
&+ \nabla_{\theta} \log p(\theta), 
\end{array}
\end{equation}
where $z$ are latent stochastic variables of the simulator. In a slight abuse of language, we will refer to $\nabla_{\theta} \log p(\theta|\text{x},z)$ as the \textit{joint score}, but it should be noted that the conventional definition of the score function in statistics (also adopted in \citet{mining_gold}) is the gradient of the log likelihood $\nabla_{\theta} \log p(\text{x}|\theta,z)$.

We can then define a direct score matching loss
\begin{equation}
\label{eq:smloss}
    \mathcal{L}_{\rm SM} = \mathbb{E}_{p(\text{x},z,\theta)} \left[ \parallel \nabla_{\theta} \log p(\theta | \text{x}, z)  - \nabla_\theta \log p_\varphi(\theta | \text{x}) \parallel_2^2 \right].
\end{equation}
Inspired by \citet{mining_gold}, this quantity is minimized by
\begin{equation}
\begin{array}{l}
     \mathbb{E}_{p(z|\text{x},\theta)} \left[ \nabla_\theta \log p(\theta |\text{x}, z) \right] \\
    ~~~~ = \mathbb{E}_{p(z|\text{x},\theta)} \left[ \nabla_\theta \log \frac{p(\theta,z | \text{x})p(\text{x})}{p(z, \text{x})} \right] \\
    ~~~~ = \mathbb{E}_{p(z|\text{x},\theta)} \left[ \nabla_\theta \log p(\theta,z | \text{x}) \right]\\
    ~~~~ = \nabla_\theta \log p(\theta| \text{x})
\end{array}    
\end{equation}

meaning that by minimizing $\mathcal{L}_{\rm SM}$ we approximate the intractable marginal score. 

Ultimately, we train our NDE using the following combined loss: 
\begin{equation}
\label{eq:ndedsloss}
    \mathcal{L} = \mathcal{L}_{\rm NLL} + \lambda \mathcal{L}_{\rm SM} \;,
\end{equation}
where $\lambda$ is a hyper-parameter which may be used to tune the score contribution to the loss function. The optimal value for $\lambda$ typically depends on the problem considered.

Note that in order to train the NDE by constraining its score $\nabla_\theta \log p_\varphi(\theta | \text{x})$, it needs to be sufficiently smooth with respect to $\theta$, motivating the development of dedicated NF architectures described in the next section.

\begin{figure*}[!h]
    \centering
    \begin{subfigure}{}
        \includegraphics[width=0.25\textwidth]{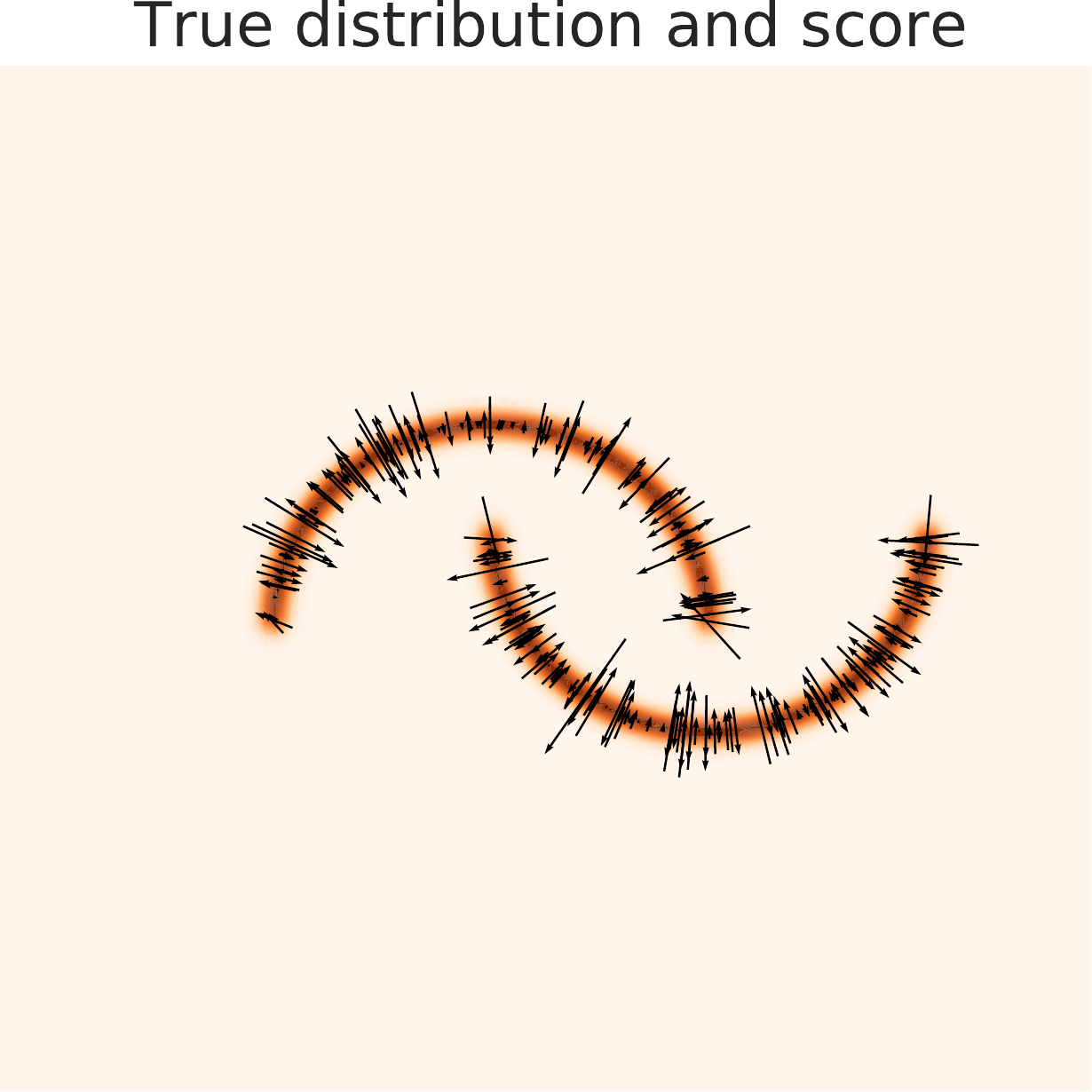}
    \end{subfigure}
    \begin{subfigure}{}
        \includegraphics[width=0.25\textwidth]{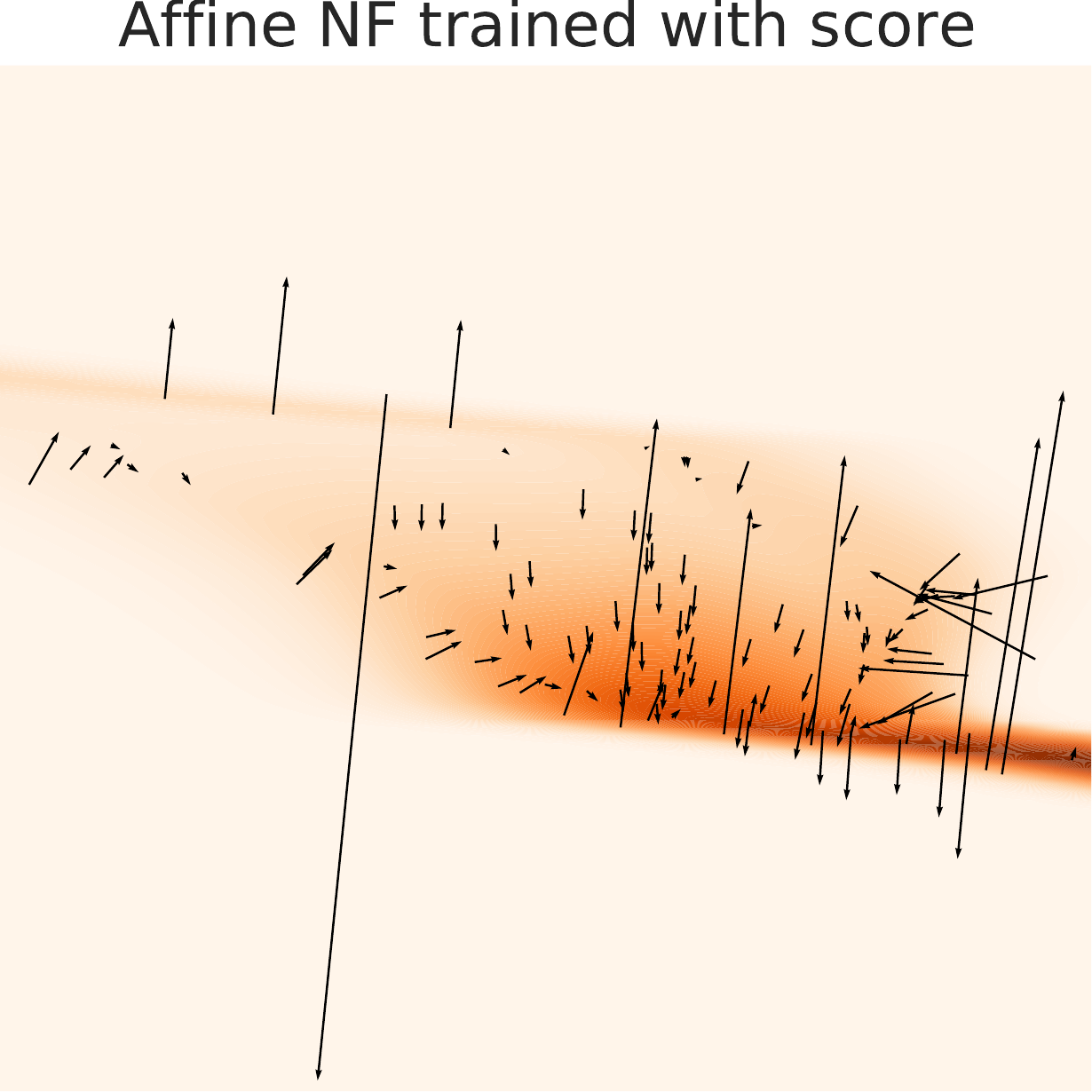}
    \end{subfigure}
    \begin{subfigure}{}
        \includegraphics[width=0.25\textwidth]{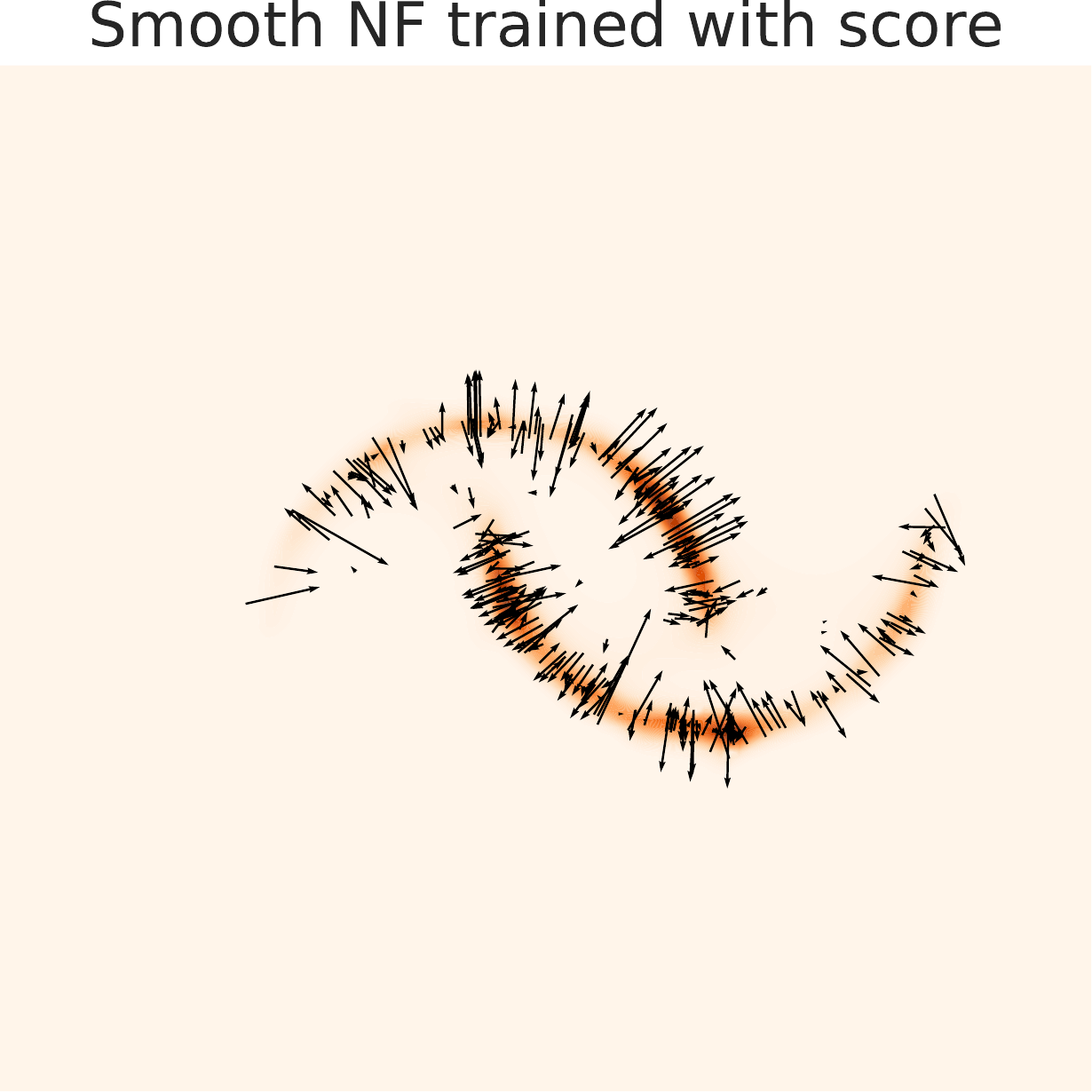}
    \end{subfigure}
    \caption{Two-moons experiment: learning a bi-modal distribution using score matching loss only.  Left: Two-moons true distribution (orange) and score (arrows). Middle: distribution and score approximated with a standard affine RealNVP \citep{realnvp2}. Right: distribution and score approximated with a smooth NF \citep{smooth_nf}. It illustrates that Affine coupling layers cannot be trained using a score matching loss.}
    \label{fig:smooth_vs_affine}
\end{figure*}

While in this work we focus on NPE, which allows for amortized inference, as demonstrated in \citet{mining_gold} similar equations can be formulated for enhancing likelihood and likelihood ratio estimation with gradient information.

\section{Smooth Normalizing Flows}
\label{smooth_nf}
To learn the approximated distribution $p_{\varphi} (\theta | \text{x})$ we focus on
NFs \citep{nfs}, a class of density estimators parameterized by neural networks that provides tractable density estimation. The key idea of NFs is to transform a simple density distribution (e.g. a multivariate Normal distribution) through a series of bijective functions to reconstruct a complex target distribution. Several strategies exist to define these bijections, here we focus on models relying on coupling transforms \citep{nice, realnvp2} $\phi: \mathbb{R}^D \mapsto \mathbb{R}^D$ of the form:
\begin{equation}
\left\{
    \begin{array}{ll}
    x_{1:d} &= z_{1:d} \\
    x_{d+1:D} &= g_{z_{1:d}}^{\varphi}(z_{d+1:D}) \label{eq:smoothnf}
    \end{array}
\right.
\end{equation}
where $g^\varphi_{z_{1:d}}$ is an invertible \textit{coupling function}, which is tuned based on the first $d$ dimensions of the input vector $z$. Standard models like the RealNVP \citep{realnvp2} use affine coupling functions $g^\varphi_y(z) = \sigma_{\varphi}(y) \cdot z + \mu_{\varphi}(y)$ with $\sigma_\varphi$ and $\mu_\varphi$ are neural networks, and $\sigma_\varphi$ strictly positive.

To be trained on both simulations and gradients using the combined loss \eqref{eq:ndedsloss}, a NF requires expressive derivatives $\nabla_\theta \log p_\varphi(\theta | \text{x})$ \eqref{eq:smloss}. If the coupling function is not smooth enough, i.e. does not have well defined and non-trivial higher-order gradients, the NF cannot be trained by score matching. We illustrate this in the case of the affine coupling of a RealNVP on the middle panel of Figure \ref{fig:smooth_vs_affine}, which fails to train under a score matching loss by lack of expressivity.

To overcome this issue, \citet{smooth_nf} propose a coupling function based on a $C^{\infty}$ diffeomorphism on $]0,1[$ 
\begin{align}
    f_{y}^\varphi(x) &= \frac{\sigma(x) - \sigma(x_0)}{\sigma(x_1) - \sigma(x_0)}\cdot (1-c_\varphi(y)) +c_\varphi(y)\cdot x \label{eq:smoothnf} \\
    \sigma(x) &= c_\varphi(y)\cdot x +  \frac{1-c_\varphi(y)}{1+\exp({-\rho(x)})}\\
    \rho(x) &= a_\varphi(y) \cdot \left(\log\left(\frac{x}{1-x}\right) + b_\varphi(y)\right)
\end{align}
with $[x_0,x_1] \subset ]0,1[$, and $a_\varphi$, $b_\varphi$, $c_\varphi$ learned using a smooth Neural Network (NN).

A single sigmoidal function like the one proposed here is not very expressive, but the sum of $C^{\infty}$-diffeomorphisms being a $C^{\infty}$-diffeomorphism, one can combine this transformation into a mixture to model a more complex bijection on $[0,1]$ (see Figure \ref{fig:smooth_transformation} in appendix). 
Because the inverse of mixtures of sigmoidal functions is not analytically defined, following \citet{smooth_nf} we compute the inverse of this coupling function numerically by the Newton-Raphson algorithm, and gradients through the inverse are obtained by the implicit function theorem \cite{implicit_dif}.

Figure \ref{fig:smooth_vs_affine} shows that using the smooth coupling function defined in \eqref{eq:smoothnf} (right plot) it is possible to train the NF by penalizing its score through the score matching loss \eqref{eq:smloss}, which was not possible for an affine coupling function (middle plot).

\section{Experiments}
\label{experiments}
We tested our method using the combined loss \eqref{eq:ndedsloss} on two different tasks: 

\textbf{Two-moons}: As a first illustration of the benefit of having access to the score to better constrain a distribution from a small number of samples in a non-conditional case, we consider a classical toy model called two-moons that consists a simple 2D bi-modal distribution. We are interested in learning the two-moons distribution $p(\text{x})$ from samples $\text{x} \sim p(\text{x})$ in two different cases: with simulations x only, and with simulations and  score $\nabla_{\text{x}}p(\text{x})$. 

We use $3$ coupling layers in our NF. The transformation parameters  $a$, $b$, $c$, defined in \eqref{eq:smoothnf}, are learned using a neural network with $4$ hidden layers of $128$ units each and $sin$ activation functions.
We trained our NF for six different training sample simulation budgets \{$20, 50, 100, 200, 500, 1000$\} and for each simulation budget we ran 10 realizations to compute the epistemic uncertainty. We evaluated the quality of the approximated distribution with the negative log-likelihood.
\begin{figure}[H]\centering
     \includegraphics[width=0.4\textwidth]{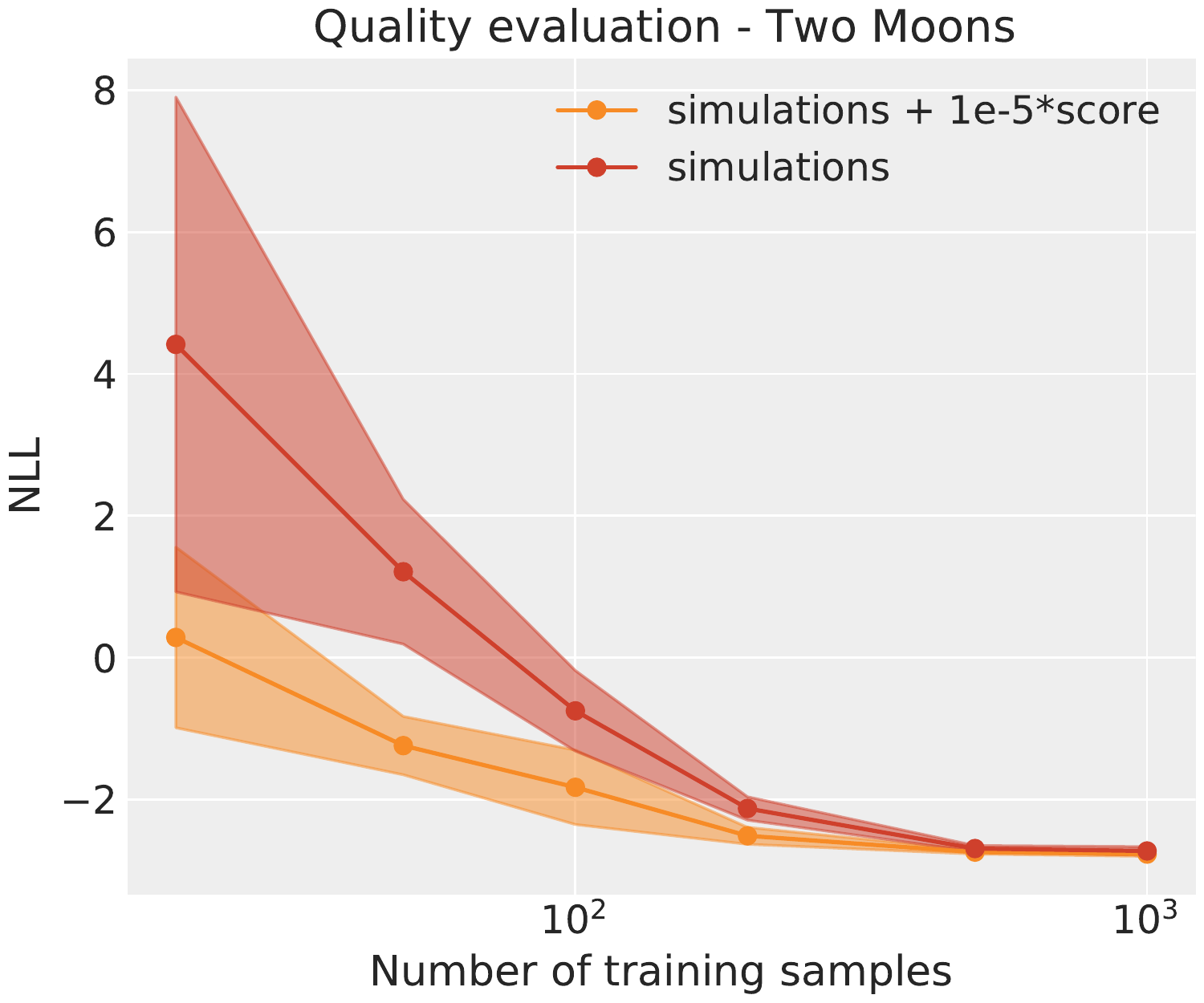}
     \caption{Two-moons inference: evolution of the quality of the distribution approximation with and without score information, measured by the negative log-likelihood, as a function of the size of the training set.}
\label{fig:inference_twomoons}
\end{figure}
We find that in this simple case the gradients help to significantly accelerate the convergence of the density estimation. As shown on Figure \ref{fig:inference_twomoons}, the approximated distribution obtained using simulation and score converge with only 200 simulations while 500 are needed for simulations alone. A comparison of the learned distributions as a function of the training set size is available in appendix in \autoref{fig:two_moons_comparison}.

\begin{figure*}\centering
     \includegraphics[width=1\textwidth]{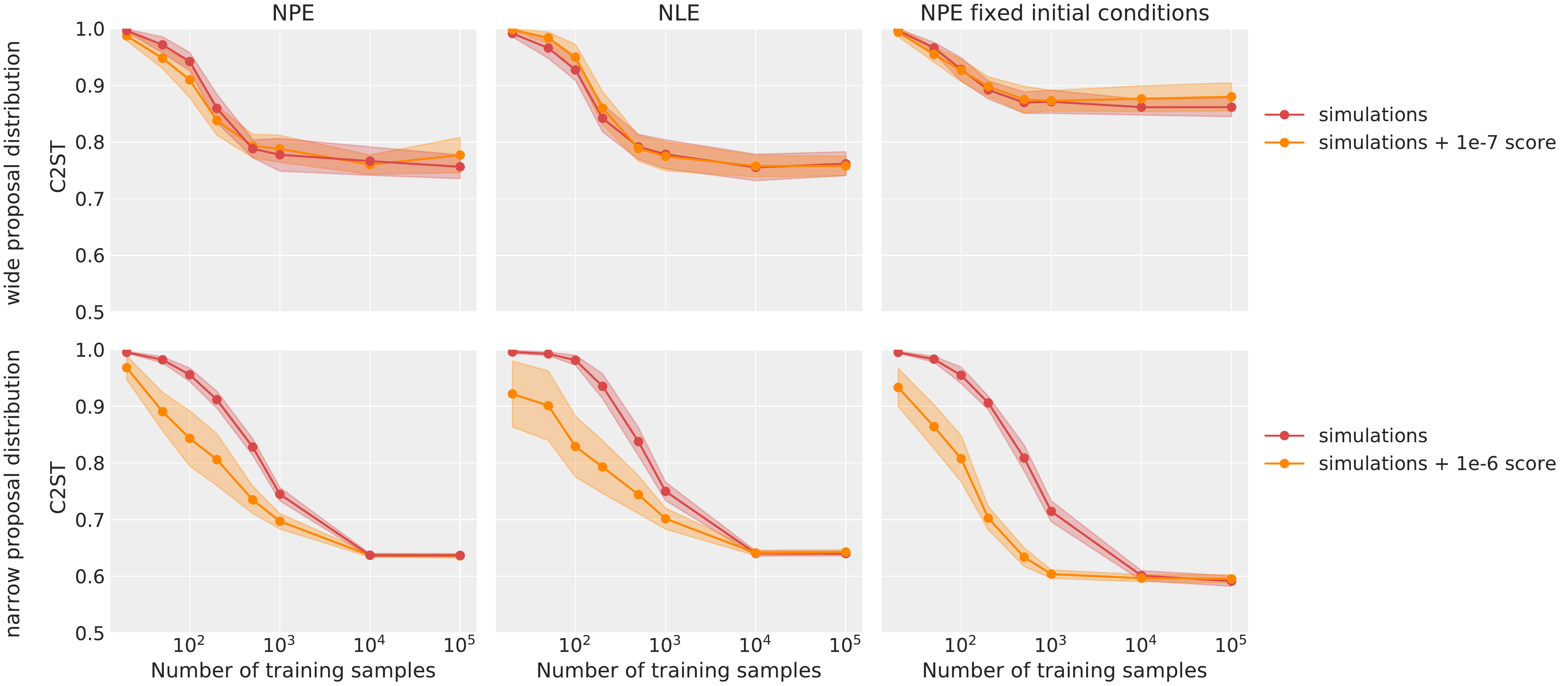}
     
     \caption{
     Lotka Volterra inference: 
     quality of the posterior approximation (C2ST metric) as a function of the number of training samples used, for three methods with (orange) and without the score (red).
     Experiments from the top row have used a wide proposal distribution \eqref{eq:widep}, while the one from the bottom row have used a narrow proposal distribution \eqref{eq:narrowp}.
     First column: NPE.
     Second column: NLE.
     Last column: NPE for Lotka Volterra task with fixed initial conditions $(x_0, Y_0) = (10,1)$.
     }
     
\label{fig:inference_lv}
\end{figure*}
\textbf{Lotka-Volterra}: 
We now consider the case of \textit{conditional} density estimation with the classical Lotka-Volterra simulation model, which consists of a system of two non-linear differential equations. It describes the interaction between two different species, the preys $X$ and the predators $Y$: 
\begin{equation}
    \begin{array}{ll}
    \frac{d X}{dt} &= \alpha  X - \beta X Y;\\[3pt]
    \frac{d Y}{dt} &= - \gamma Y + \delta X Y,
    \end{array}
\end{equation}
with, $\alpha,\beta,\gamma$ and $\delta$ the inferred parameters  $\theta \in \mathbb{R}^4$.

The simulations are  20 statistics that correspond to the number of each population $X$ and $Y$ at $10$ different times. Hence our simulations correspond to $\text{x} = (\text{x}_1,..,\text{x}_{10},\text{y}_1,..,\text{y}_{10}) \in \mathbb{R}^{20}$.
We considered the initial conditions as our latent variables $z$ with prior
\begin{equation}
    \begin{array}{ll}
z =  \begin{bmatrix}
\text{x}_0\\ \text{y}_0
\end{bmatrix} \sim  \text{LogNormal}\left(\log(3),0.5\right),
    \end{array}
\end{equation}
which lead to the following simulations
\begin{equation}
    \begin{array}{ll}
\text{x}|\theta = 
\begin{bmatrix}
\text{x}_1, & .. , & \text{x}_{10}\\
\text{y}_1, & .. , & \text{y}_{10}
\end{bmatrix}
\sim \text{LogNormal}\left(\log
\begin{bmatrix}
X\\
Y
\end{bmatrix}
,0.1\right).
    \end{array}
\end{equation}

We expect that the quality of the posterior approximation $p_{\varphi}(\theta|x)$ no longer depends only on capturing the shape of the distribution, but also on successfully conditioning it on observations. To illustrate these two factors, we design two experiments, one with a wide proposal distribution (compare to the posterior)
\begin{equation}
\label{eq:widep}
    \begin{array}{ll}
    \begin{bmatrix}
\alpha\\ \beta\\ \gamma\\ \delta
\end{bmatrix}
\sim \text{LogNormal}\left(    \begin{bmatrix}
-0.125\\ -3 \\ -0.125\\ -3
\end{bmatrix},0.5\right), \\
    \end{array}
\end{equation}
for which we expect errors from the conditioning to dominate, and a narrow proposal distribution,
\begin{equation}
\label{eq:narrowp}
    \begin{array}{ll}
    \begin{bmatrix}
\alpha\\ \beta\\ \gamma\\ \delta
\end{bmatrix}
\sim \text{LogNormal}\left(    \begin{bmatrix}
-0.125\\ -3 \\ -0.125\\ -3
\end{bmatrix},0.03\right), \\
    \end{array}
\end{equation}
for which the main task is to focus on the posterior shape rather than finding its location.

As a first step, it is standard practice in SBI to compress x using an external neural network into a summary statistic of same dimensionality as the number of parameters. Moreover, this will allow us to make a fairer comparison between Neural Likelihood Estimation (NLE) and our NPE approach since in both cases the density estimation problem will be of same dimensionality and thus the same NF architecture can be used. In our case,
 we compressed x using a 1D CNN $r_\phi$ trained externally. We trained the CNN under a NLL loss into a low dimensional $4-d$ summary statistics $y$, and used $4$ smooth transformations in our NF. The transformation parameters  $a$, $b$, $c$, defined in \eqref{eq:smoothnf}, were learned using  a small NN (enough due to the external compressor mentioned above): $2$ hidden layers of $128$ units and $silu$ activation functions \cite{silu}. To make our NDE conditional on $y$, this variable is used as an additional input to the coupling layers $x$. This yields  the approximated distribution $p_{\varphi} (\theta | y = r_\phi(\text{x}))$.\\
 
We compared our method (NPE with score) with SCANDAL \citep[NLE with score from][]{mining_gold}. We used the same compressor and NF architecture and run a MCMC to get the approximated posterior from the learned likelihood $p_{\varphi} (y = r_\phi(\text{x})
|\theta )$.

For this task, we trained the NFs for eight different training sample simulations budget
\{20, 50, 100, 200, 500, $10^{3}$, $10^{4}$, $10^{5}$\} and for each simulation budget we ran 20 realizations to compute the epistemic uncertainty. We sample these simulations according to two different proposal distributions described above. We evaluated the quality of the approximated posterior with the Classifier $2$-Sample Tests (C2ST) metric \citep{sbibm}.  If the two samples come from the same distribution the classifier's prediction is $0.5$.

The two figures on the left in the first row of Figure  \ref{fig:inference_lv} show that, in the case of a wide proposal distribution, adding the gradients in the inference process do not help to constrain the posterior. The score information does not help to constrain the conditioning of the density, i.e. to help constrain the overall location of the posterior mass. 
However, in the case of a narrow proposal distribution, the two figures on the left in the second row of the Figure \ref{fig:inference_lv}, we find that score information significantly helps accelerating inference, for both NPE and NLE. This can be understood as in this case the location of the posterior mass is already well defined, and the score information contributes to constraining the shape of the posterior.

Finally, we highlight the impact of the stochastic score $\nabla_{\theta} \log p(\theta|\text{x},z)$ by repeating the same experiment with fixed initial conditions $(X_0, Y_0) = (10,1)$. We found that the gradients help more without latent variables in the case of a narrow proposal distribution (see the last figures in the second row of Figure \ref{fig:inference_lv}). This result is expected as the score in the fixed latent variable case is no longer stochastic, and can directly help to constrain the posterior distribution. However, fixed initial conditions make the posterior narrower and thus in the case of wide proposal distribution (see the last figures in the first row of Figure \ref{fig:inference_lv}) the gradients help even less.

\section{Discussion}
\label{discussion}

To train the NF using simulations and score, we used a smooth NF from \citet{smooth_nf}. Other types of differentiable $C^{\infty}$ NF, such as ODE Flows \citet{ffjord}, could be used in practice.\\
We demonstrated that having access to gradients from the simulation is only beneficial for the density shape estimation. This points to a high interest of having access to score information to refine rough estimates, and thus to sequential density estimation methods.
We expect these tools to become increasingly useful in astrophysics with the advent of automatically differentiable physical simulators like the FlowPM cosmological simulation code \citep{flowpm}.

\nocite{langley00}

\bibliography{example_paper}

\begin{thebibliography}{21}
\providecommand{\natexlab}[1]{#1}
\providecommand{\url}[1]{\texttt{#1}}
\expandafter\ifx\csname urlstyle\endcsname\relax
  \providecommand{\doi}[1]{doi: #1}\else
  \providecommand{\doi}{doi: \begingroup \urlstyle{rm}\Url}\fi

\bibitem[Blondel et~al.(2021)Blondel, Berthet, Cuturi, Frostig, Hoyer,
  Llinares-López, Pedregosa, and Vert]{implicit_dif}
Blondel, M., Berthet, Q., Cuturi, M., Frostig, R., Hoyer, S., Llinares-López,
  F., Pedregosa, F., and Vert, J.-P.
\newblock Efficient and modular implicit differentiation, 2021.
\newblock URL \url{https://arxiv.org/abs/2105.15183}.

\bibitem[Blum \& Fran{\c{c}}ois(2009)Blum and Fran{\c{c}}ois]{pe}
Blum, M. G.~B. and Fran{\c{c}}ois, O.
\newblock Non-linear regression models for approximate bayesian computation.
\newblock \emph{Statistics and Computing}, 20\penalty0 (1):\penalty0 63--73,
  mar 2009.
\newblock \doi{10.1007/s11222-009-9116-0}.
\newblock URL \url{https://doi.org/10.10072Fs11222-009-9116-0}.

\bibitem[Brehmer et~al.(2020)Brehmer, Louppe, Pavez, and Cranmer]{mining_gold}
Brehmer, J., Louppe, G., Pavez, J., and Cranmer, K.
\newblock Mining gold from implicit models to improve likelihood-free
  inference.
\newblock \emph{Proceedings of the National Academy of Sciences}, 117\penalty0
  (10):\penalty0 5242--5249, feb 2020.
\newblock \doi{10.1073/pnas.1915980117}.
\newblock URL \url{https://doi.org/10.1073%2Fpnas.1915980117}.

\bibitem[Cranmer et~al.(2015)Cranmer, Pavez, and Louppe]{lr1}
Cranmer, K., Pavez, J., and Louppe, G.
\newblock Approximating likelihood ratios with calibrated discriminative
  classifiers, 2015.
\newblock URL \url{https://arxiv.org/abs/1506.02169}.

\bibitem[Dinh et~al.(2014)Dinh, Krueger, and Bengio]{nice}
Dinh, L., Krueger, D., and Bengio, Y.
\newblock Nice: Non-linear independent components estimation, 2014.
\newblock URL \url{https://arxiv.org/abs/1410.8516}.

\bibitem[Dinh et~al.(2016)Dinh, Sohl-Dickstein, and Bengio]{realnvp2}
Dinh, L., Sohl-Dickstein, J., and Bengio, S.
\newblock Density estimation using real nvp, 2016.
\newblock URL \url{https://arxiv.org/abs/1605.08803}.

\bibitem[Durkan et~al.(2020)Durkan, Murray, and Papamakarios]{lr4}
Durkan, C., Murray, I., and Papamakarios, G.
\newblock On contrastive learning for likelihood-free inference, 2020.
\newblock URL \url{https://arxiv.org/abs/2002.03712}.

\bibitem[Elfwing et~al.(2017)Elfwing, Uchibe, and Doya]{silu}
Elfwing, S., Uchibe, E., and Doya, K.
\newblock Sigmoid-weighted linear units for neural network function
  approximation in reinforcement learning, 2017.

\bibitem[Grathwohl et~al.(2018)Grathwohl, Chen, Bettencourt, Sutskever, and
  Duvenaud]{ffjord}
Grathwohl, W., Chen, R. T.~Q., Bettencourt, J., Sutskever, I., and Duvenaud, D.
\newblock Ffjord: Free-form continuous dynamics for scalable reversible
  generative models, 2018.
\newblock URL \url{https://arxiv.org/abs/1810.01367}.

\bibitem[Greenberg et~al.(2019)Greenberg, Nonnenmacher, and Macke]{npe3}
Greenberg, D.~S., Nonnenmacher, M., and Macke, J.~H.
\newblock Automatic posterior transformation for likelihood-free inference,
  2019.

\bibitem[Izbicki et~al.(2014)Izbicki, Lee, and Schafer]{lr2}
Izbicki, R., Lee, A.~B., and Schafer, C.~M.
\newblock High-dimensional density ratio estimation with extensions to
  approximate likelihood computation.
\newblock 2014.
\newblock \doi{10.48550/ARXIV.1404.7063}.
\newblock URL \url{https://arxiv.org/abs/1404.7063}.

\bibitem[Köhler et~al.(2021)Köhler, Krämer, and Noé]{smooth_nf}
Köhler, J., Krämer, A., and Noé, F.
\newblock Smooth normalizing flows, 2021.
\newblock URL \url{https://arxiv.org/abs/2110.00351}.

\bibitem[Lueckmann et~al.(2017)Lueckmann, Goncalves, Bassetto, Öcal,
  Nonnenmacher, and Macke]{npe2}
Lueckmann, J.-M., Goncalves, P.~J., Bassetto, G., Öcal, K., Nonnenmacher, M.,
  and Macke, J.~H.
\newblock Flexible statistical inference for mechanistic models of neural
  dynamics, 2017.

\bibitem[Lueckmann et~al.(2018)Lueckmann, Bassetto, Karaletsos, and
  Macke]{nle2}
Lueckmann, J.-M., Bassetto, G., Karaletsos, T., and Macke, J.~H.
\newblock Likelihood-free inference with emulator networks.
\newblock 2018.
\newblock \doi{10.48550/ARXIV.1805.09294}.
\newblock URL \url{https://arxiv.org/abs/1805.09294}.

\bibitem[Lueckmann et~al.(2021)Lueckmann, Boelts, Greenberg, Gonçalves, and
  Macke]{sbibm}
Lueckmann, J.-M., Boelts, J., Greenberg, D.~S., Gonçalves, P.~J., and Macke,
  J.~H.
\newblock Benchmarking simulation-based inference, 2021.
\newblock URL \url{https://arxiv.org/abs/2101.04653}.

\bibitem[Modi et~al.(2020)Modi, Lanusse, and Seljak]{flowpm}
Modi, C., Lanusse, F., and Seljak, U.
\newblock Flowpm: Distributed tensorflow implementation of the fastpm
  cosmological n-body solver, 2020.
\newblock URL \url{https://arxiv.org/abs/2010.11847}.

\bibitem[Papamakarios \& Murray(2018)Papamakarios and Murray]{npe1}
Papamakarios, G. and Murray, I.
\newblock Fast $\epsilon$-free inference of simulation models with bayesian
  conditional density estimation, 2018.

\bibitem[Papamakarios et~al.(2018)Papamakarios, Sterratt, and Murray]{nle1}
Papamakarios, G., Sterratt, D.~C., and Murray, I.
\newblock Sequential neural likelihood: Fast likelihood-free inference with
  autoregressive flows, 2018.
\newblock URL \url{https://arxiv.org/abs/1805.07226}.

\bibitem[Rezende \& Mohamed(2015)Rezende and Mohamed]{nfs}
Rezende, D.~J. and Mohamed, S.
\newblock Variational inference with normalizing flows, 2015.
\newblock URL \url{https://arxiv.org/abs/1505.05770}.

\bibitem[Thomas et~al.(2016)Thomas, Dutta, Corander, Kaski, and Gutmann]{lr3}
Thomas, O., Dutta, R., Corander, J., Kaski, S., and Gutmann, M.~U.
\newblock Likelihood-free inference by ratio estimation, 2016.
\newblock URL \url{https://arxiv.org/abs/1611.10242}.

\bibitem[Wood(2010)]{wood_statistical_2010}
Wood, S.~N.
\newblock Statistical inference for noisy nonlinear ecological dynamic systems.
\newblock \emph{Nature}, 466\penalty0 (7310):\penalty0 1102--1104, August 2010.
\newblock ISSN 1476-4687.
\newblock \doi{10.1038/nature09319}.
\newblock URL \url{https://doi.org/10.1038/nature09319}.

\end{thebibliography}
\bibliographystyle{icml2022}

\newpage
\appendix
\onecolumn
\section{Additional plots}

\begin{figure}[!h]\centering
     \includegraphics[width=0.4\textwidth]{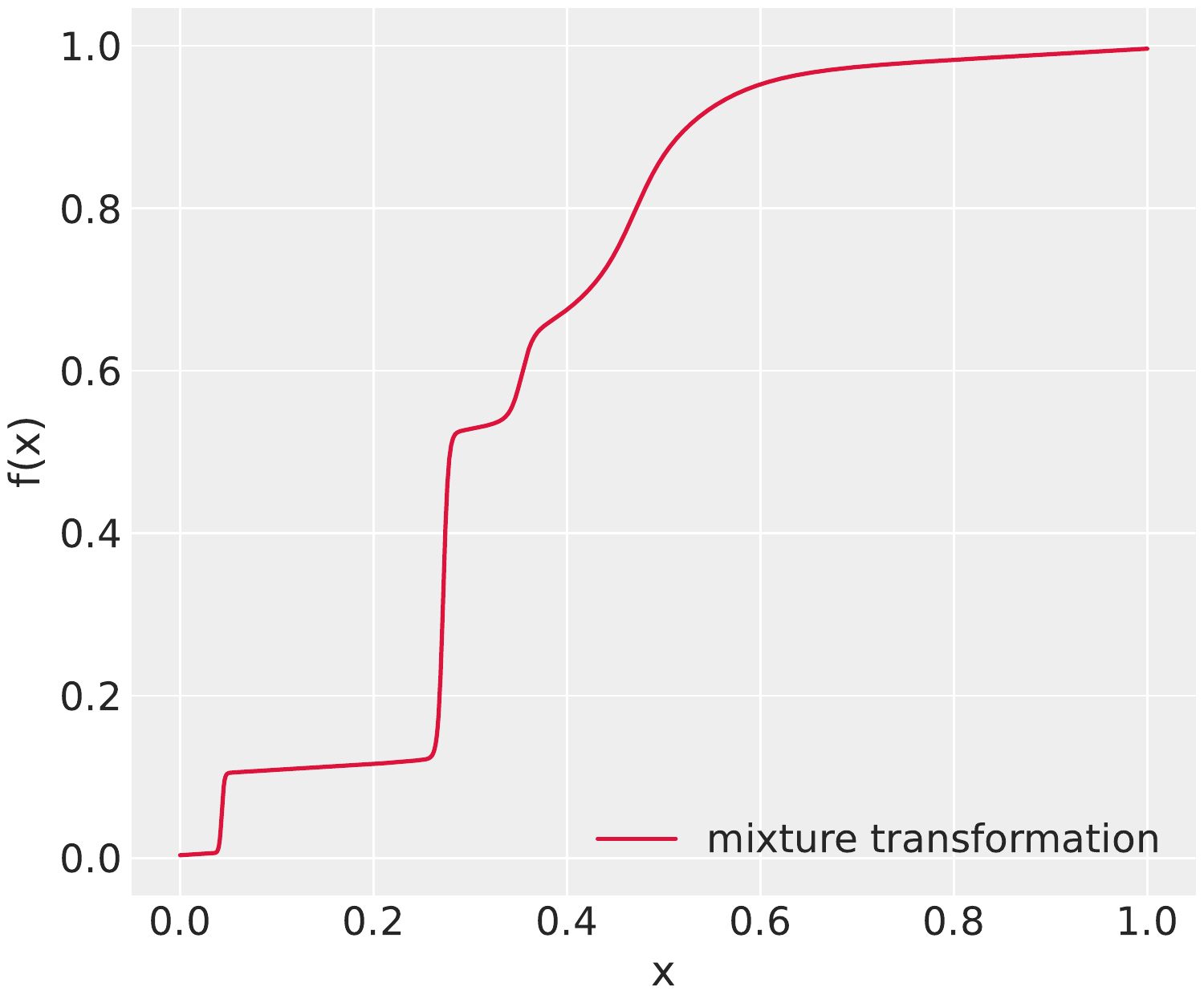}
     \caption{Mixture of 5 smooth transformations used to create a smooth Normalizing Flow.}
\label{fig:smooth_transformation}
\end{figure}

\begin{figure*}[!h]
    \centering
    \begin{subfigure}{}
        \includegraphics[width=0.4\textwidth]{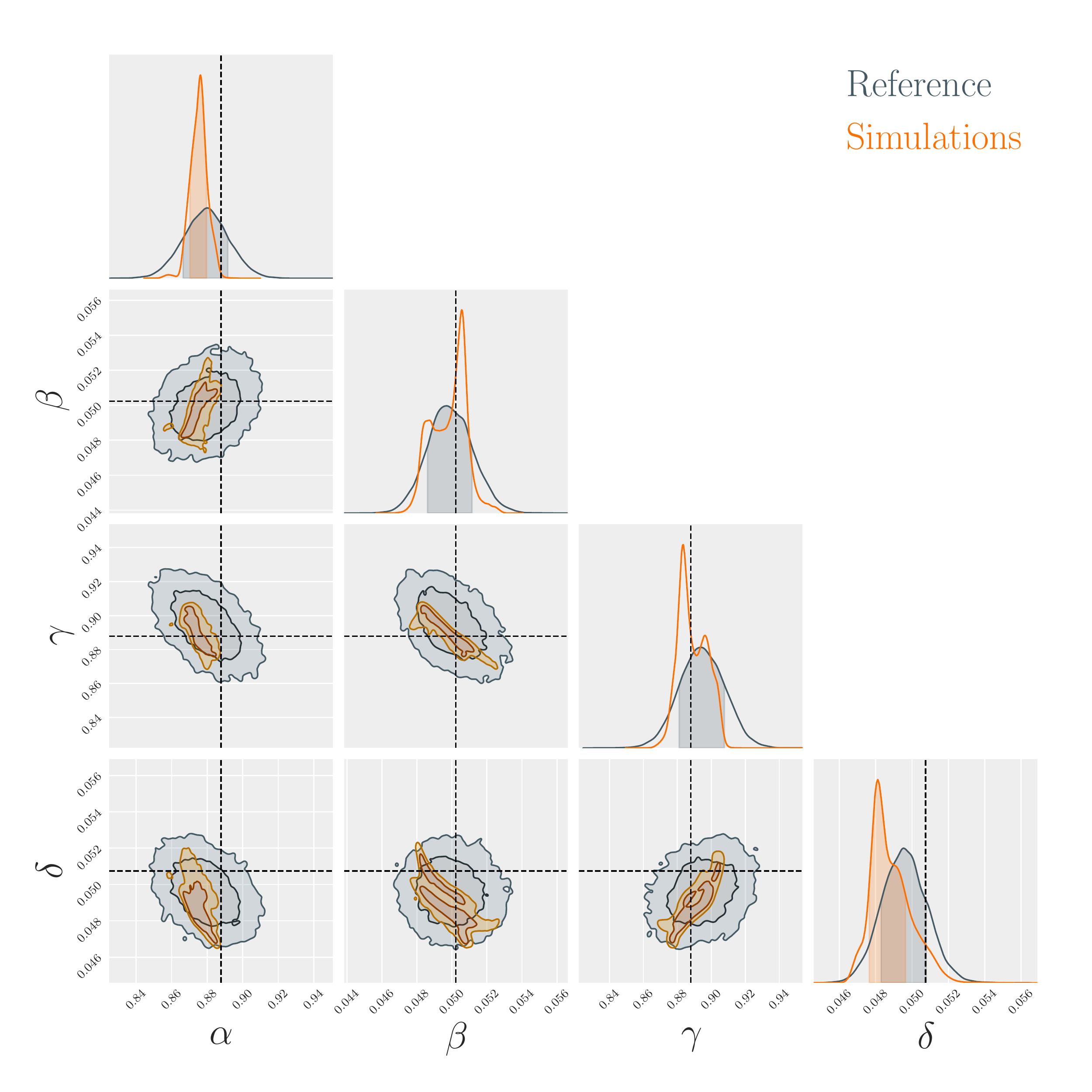}
    \end{subfigure}
    \begin{subfigure}{}
        \includegraphics[width=0.4\textwidth]{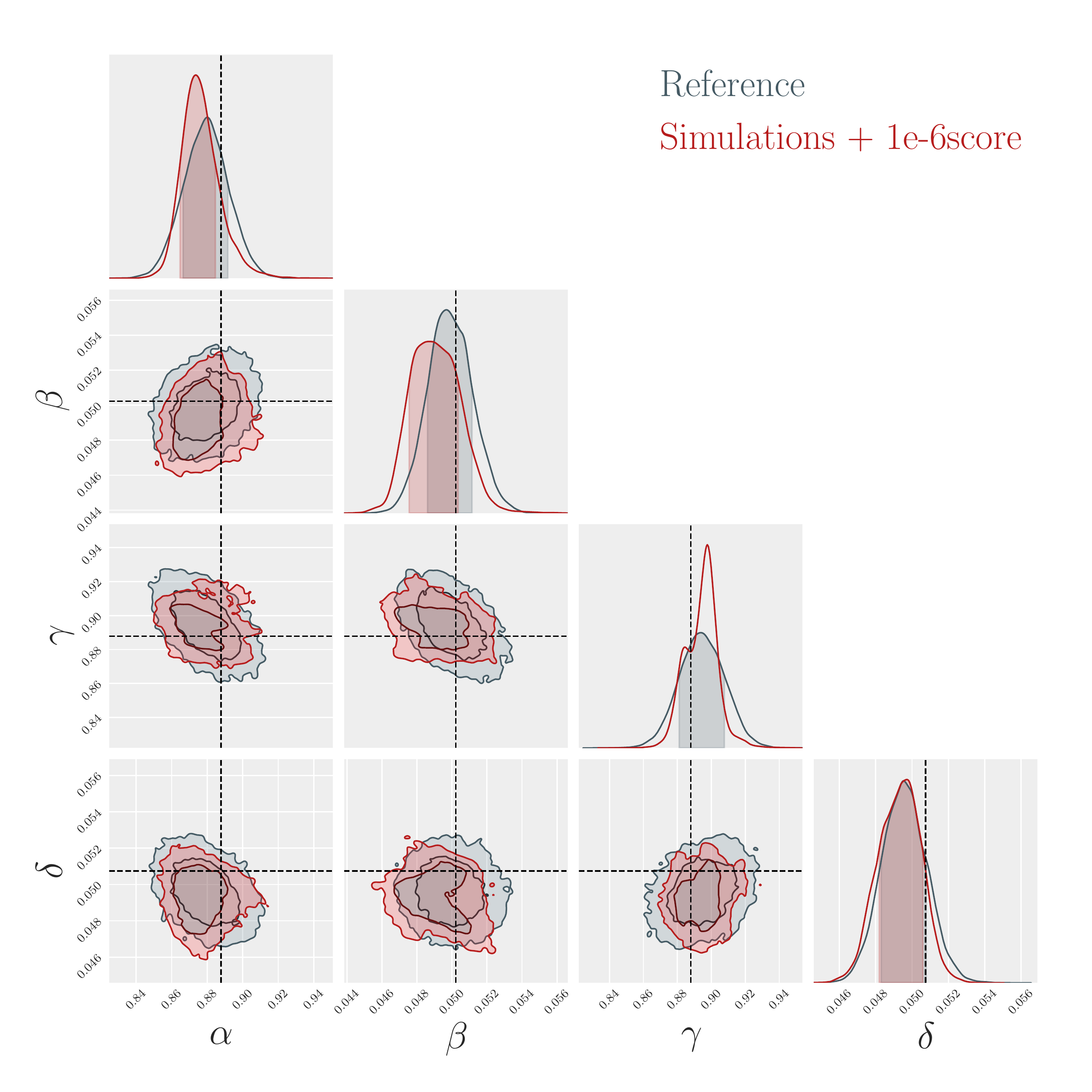}
    \end{subfigure}
    \caption{Lotka Volterra posterior comparison between the reference posterior (blue), the posterior obtained using only simulations (orange on the left) and the one obtained using both simulations and gradients (red on the right). These posteriors are approximated using NPE method with only 50 simulations in the case of a narrow proposal distribution.}
    \label{fig:contourplot_lv}
\end{figure*}

\begin{figure*}[!h]
    \centering
    \begin{subfigure}{}
        \includegraphics[width=0.17\textwidth]{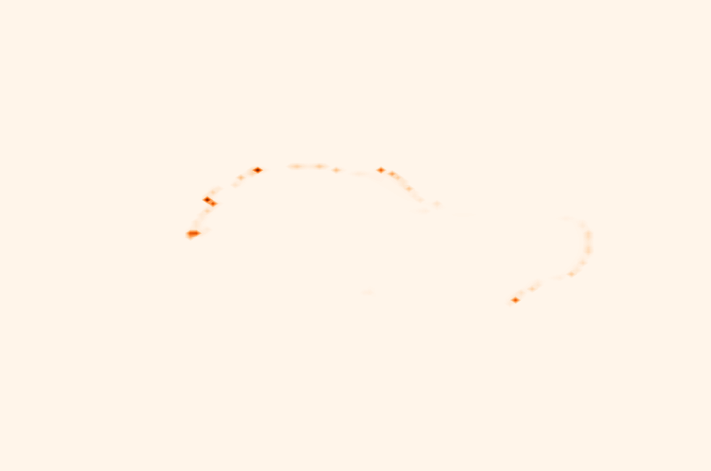}
    \end{subfigure}
    \begin{subfigure}{}
        \includegraphics[width=0.17\textwidth]{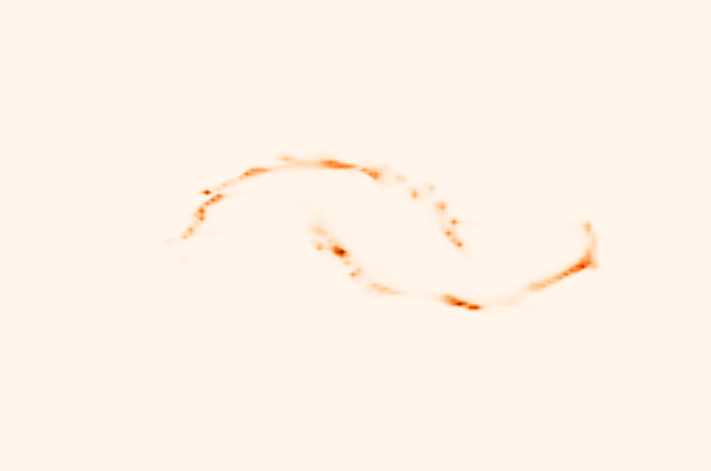}
    \end{subfigure}
    \begin{subfigure}{}
        \includegraphics[width=0.17\textwidth]{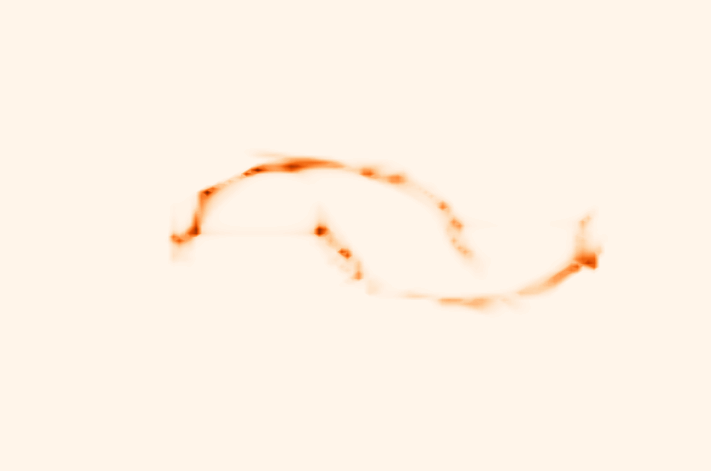}
    \end{subfigure}
    \begin{subfigure}{}
        \includegraphics[width=0.17\textwidth]{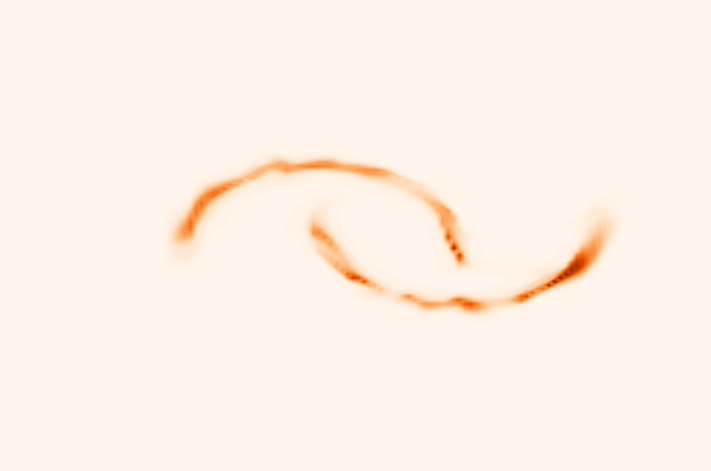}
    \end{subfigure}
    \begin{subfigure}{}
        \includegraphics[width=0.17\textwidth]{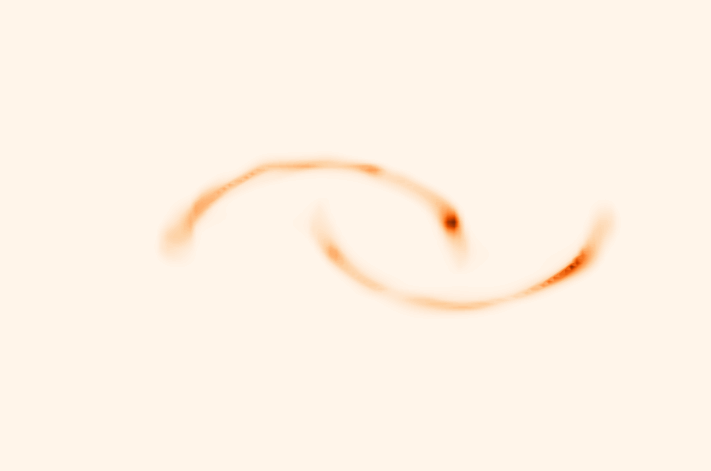}
    \end{subfigure}
    \begin{subfigure}{}
        \includegraphics[width=0.17\textwidth]{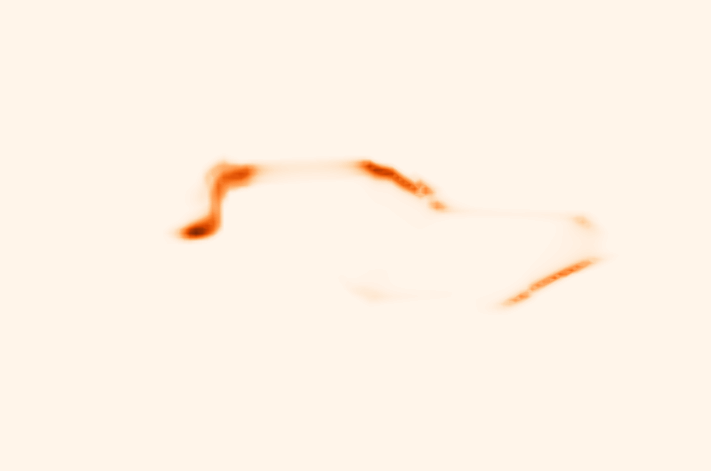}
    \end{subfigure}
    \begin{subfigure}{}
        \includegraphics[width=0.17\textwidth]{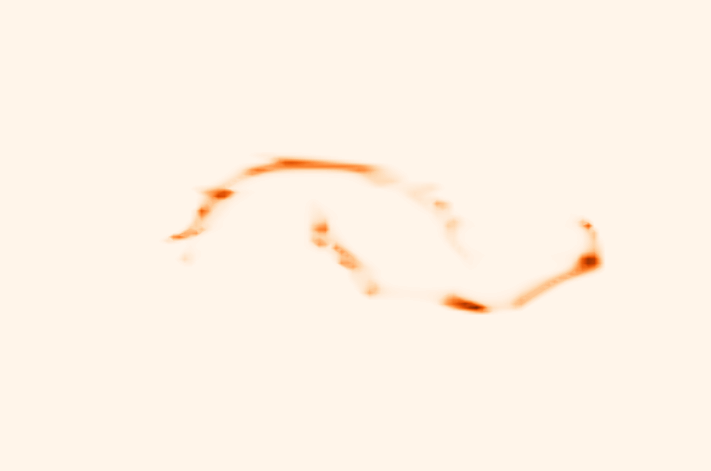}
    \end{subfigure}
    \begin{subfigure}{}
        \includegraphics[width=0.17\textwidth]{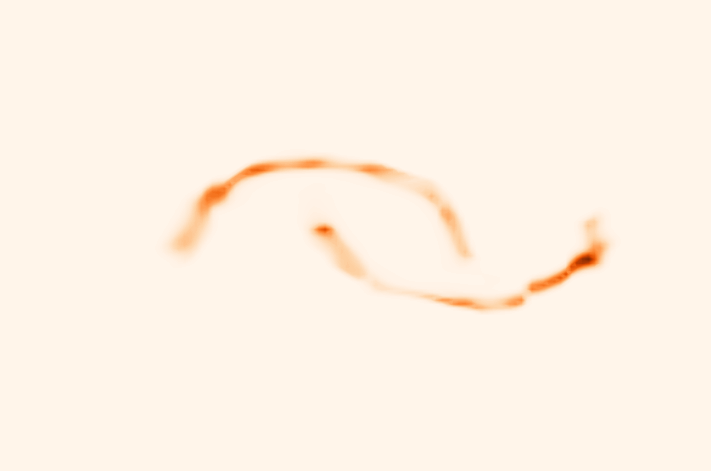}
    \end{subfigure}
    \begin{subfigure}{}
        \includegraphics[width=0.17\textwidth]{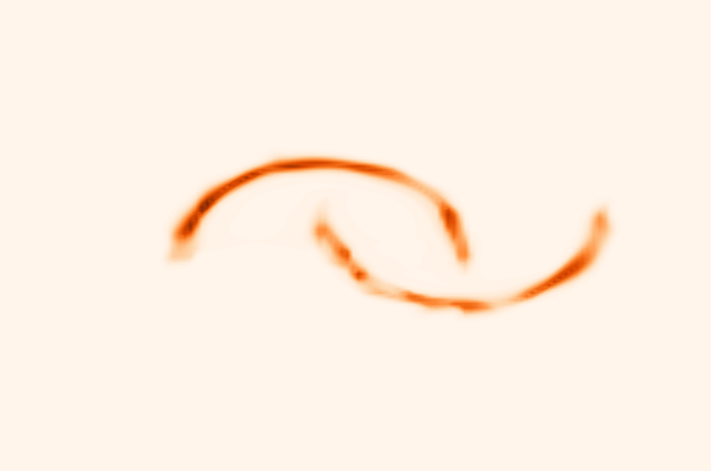}
    \end{subfigure}
    \begin{subfigure}{}
        \includegraphics[width=0.17\textwidth]{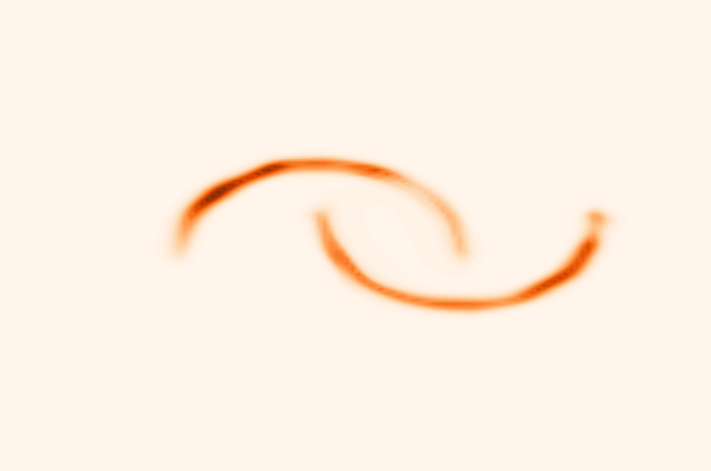}
    \end{subfigure}
    \caption{Two Moons distribution approximated using, from left to right, $20, 100, 200, 500 , 1000$ simulations. The first row correspond to the distribution learned using simulations only and the second row learned on simulations and score.}
    \label{fig:two_moons_comparison}
\end{figure*}

\end{document}